\documentclass[runningheads,a4paper]{llncs}

\usepackage{amssymb}
\usepackage{graphicx}
\usepackage{algorithm}
\usepackage{algorithmic}
\usepackage{hyperref}
\usepackage[numbers]{natbib}

\usepackage{url}
\urldef{\mailsc}\path|Djallel.Bouneffouf}@Orange.com|    
\newcommand{\keywords}[1]{\par\addvspace\baselineskip
\noindent\keywordname\enspace\ignorespaces#1}

\begin{document}

\mainmatter  

\title{Freshness-Aware Thompson Sampling}

\titlerunning{Lecture Notes in Computer Science: Authors' Instructions}

%
%
\author{Djallel Bouneffouf
\thanks{}
}
\authorrunning{Lecture Notes in Computer Science: Authors' Instructions}

\institute{Orange Labs, 2, avenue Pierre Marzin, 22307 Lannion, France,\\
\mailsc\\
}

%
%

\toctitle{Lecture Notes in Computer Science}
\tocauthor{Authors' Instructions}
\maketitle

\begin{abstract}
To follow the dynamicity of the user's content, researchers have recently started to model interactions between users and the Context-Aware Recommender Systems (CARS) as a bandit problem where the system needs to deal with exploration and exploitation dilemma. In this sense, we propose to study the freshness of the user's content in CARS through the bandit problem. We introduce in this paper an algorithm named Freshness-Aware Thompson Sampling (FA-TS) that manages the recommendation of fresh document according to the user's risk of the situation. The intensive evaluation and the detailed analysis of the experimental results reveals several important discoveries in the exploration/exploitation (exr/exp) behaviour.\keywords{CARS, Thompson Sampling, Contextual bandits }
\end{abstract}

\section{Introduction}
\label{sec:1}
Mobile technologies have made access to a huge collection of information, anywhere and any-time. In this sense, recommender systems must promptly identify the importance of documents to recommend in the great location and moment.
Recently, CARS tackle this problem by relating the user's interest to the user's situation (time, location, friends). However, they cannot avoid to recommend the same document under the same situations. As a result, a small set of documents are recommended again and again and then are seen as favourite documents, however recommend the same set of documents many times in a short period makes the users feel bored. 
Works found in literature \cite{21, 13, TSCB} tackle this problem by addressing the recommendation as a need for balancing exr/exp studied in the "bandit algorithm". Actually the greatest result in exr/exp is performed by the Thompson Sampling (TS), but its drawback is in the none consideration of the freshness of document in the recommendation.
The Freshness can be considered as the strength of strangeness or the amount of forgotten experience \cite{ebbinghaus1913memory}, and it leads the system to recommend some documents that have not been clicked for a long time because these documents are fresh to users even though they do not click to them multiple times. To this effect, we introduce in this paper an algorithm named Freshness-Aware Thompson Sampling (FATS) that achieves this goal by balancing adaptively the exr/exp trade-off according to the user's situation and the document's freshness.
This algorithm extends the TS strategy by exploring fresh documents in suitable user's situations.

The remaining of the paper is organized as follows. Section 2 reviews related works. Section 3 gives key notion used in the paper. Section 4 describes the algorithms involved in the proposed approach. The experimental evaluation is illustrated in Section 5. The last section concludes the paper and points out possible directions for future work.
\section{Related Work}
\label{sec:related}
We refer, in the following, techniques that study the different dimensions of our problem.
\textbf{Multi-Armed Bandit Problem in RS.} 
Recently, research works are dedicated to study the multi-armed bandit problem in RS, considering the user's behaviour as the context.   
In \cite{BouneffoufBG12}, authors model CARS as a contextual bandit problem. The authors propose an algorithm called Contextual-$\epsilon$-greedy which a perform recommendation sequentially recommends documents based on contextual information about the users' documents.
In \cite{TSCB}, authors analyse the TS in contextual bandit problem. The study demonstrate that it has better empirical performance compared to the state-of-art methods.  
The authors in \cite{BouneffoufBG12, TSCB} describe a smart way to balance exr/exp, but do not consider the user's context and document freshness during the recommendation.

\textbf{User's Content Dynamicity in RS.} 
To follow the dinamicity of the user's content, the authors in \cite{Mortal} formulate and study a new variant of the k-armed bandit problem, motivated by e-commerce applications. In their model, arms have (stochastic) lifetime after which they expire. In this setting an algorithm needs to continuously explore new arms, contrarily to the standard k-armed bandit model in which arms are available indefinitely and exploration is reduced once an optimal arm is identified. In this work the dynamicity of the content is considered but the authors do not address the notion of freshness.
A notion of freshness of document is used in \cite{HuEtAl2011}, where the authors propose an RS that considers the freshness of music in recommendation. However they neither consider the freshness in CARS nor in multi-armed bandit problem.
 
\textbf{The Risk-Aware Decision.}
The risk-aware decision has been studied for a long time in reinforcement learning, where the risk is defined as the reward criteria that not only takes into account the expected reward, but also some additional statistics of the total reward, such as its variance or standard deviation \cite{Luenberger1998InvestmentScience}.
In RS the risk is recently studied. The authors in \cite{BouneffoufBG13} consider the risk of the situations in the recommendation process, and the study yields to the conclusion that considering the risk level of the
situation on the exr/exp strategy significantly increases the performance of the recommender system.
 
\textbf{Contribution.} 
From this state of the art we observe that none of the existing works have studied the correlation between the user's situation risk and the freshness document recommendation. This is precisely what we intend to do with Freshness-Aware Thompson Sampling (FATS), the proposing algorithm exploits the following new features: (1) The algorithm takes into consideration the document's freshness in its exr/exp trade-off by considering the "Forgetting Curve" to assess freshness and evaluate favouredness. (2) The algorithm manages the recommendation of fresh documents according to the user's situation, where the fresh documents are more explored in non-risky situation (the user is at home the user may be interested by a freshness documents) rather than risky or critical situation (the user is at the office, in a meeting or with a client) the system has to do less exploration to avoid disturbing the user.
\section{Key Notion}
This section focuses on introducing the key notions used in this paper.

\textbf{Situation:} A situation is an external semantic interpretation of low-level context data, enabling a higher-level specification of human behaviour. More formally, a situation $S$ is a n-dimensional vector,  $S=(O_{\delta_{1}}.c_1,O_{\delta_{2}}.c_2,...,O_{\delta_{n}}.c_n)$ where each $c_i$ is a concept of an ontology $O_{\delta_{i}}$ representing a context data dimension. 
According to our need, we consider a situation as a 3-dimensional vector 
$S=(O_{Location}.c_i, O_{Time}.c_j, O_{Social}.c_k)$ where $c_i, c_j, c_k $ are concepts of Location, Time and Social ontologies.
 
\textbf{User preferences:} User preferences $UP$ are deduced during the user navigation activities.  $UP\subseteq D \times A \times V$ where $D$ is a set of documents, $A$ is a set of preference attributes and $V$ a set of values. We focus on the following preference attributes: \textit{click}, \textit{fail} , \textit{time} and \textit{recom} which respectively correspond to the number of clicks for a document, number of failure (recommended and not clicked), the time spent on a document and the number of times it was recommended.

\textbf{The user model:} The user model is structured as a case base composed of a set of situations with their corresponding $UP$, denoted $UM=\{(S^i; UP^i)\}$, where $S^i \in S$ is the user situation and
$UP^i \in UP $ its user preferences.

\textbf{Definition of risk:} "The risk in recommender systems is the possibility to disturb or to upset the user (which leads to a bad answer of the user)".

From the precedent definition of the risk, we have proposed to consider in our system Critical Situations (CS) which is a set of situations where the user needs the best information that can be recommended by the system, because he can not be disturbed. This is the case, for instance, of a professional meeting. In such a situation, the system must exclusively perform exploitation rather than exploration-oriented learning. In other cases where the risk of the situation is less important (like for example when the user is using his information system at home, or he is on holiday with friends), the system can make some exploration by recommending information without taking into account his interest.

To consider the risk level of the situation in RS, we go further in the definition of situation by adding it a risk level $R$, as well as one to each concept:
$S [R]$=($O_{\delta_{1}}.c_1 [cv_1],O_{\delta_{2}}.c_2 [cv_2],...$, $O_{\delta_{n}}.c_n [cv_n]$) where 
$CV$=$\{cv_1, cv_2,...,cv_n\}$ is the set of risk levels assigned to concepts, $cv_ i \in [0,1]$. $R\in [0,1]$ is the risk level of situation $\textit{S}$, and the set of situations with $R=1$ are considered as CS. 

\textbf{Definition (Situation Bandit Problem).}
In a situation bandits problem, there is a distribution $P$ over $(S^i, r(d_1),..., r(d_k))$, where $S$ is the situation, $ d_{i} \in D$ is one of the $k$ document to be recommended, and $r(d) \in [0,1]$ is the reward for document $d$. The problem is a repeated game: on each round, a sample $(S^i, r(d_1), ..., r(d_k))$ is drawn from $P$, the situation $S$ is announced, and then for one document chosen by the system, its reward $r(d)$ is revealed.
 
\textbf{Definition (\textbf{Thompson Sampling}).}
The Thompson Sampling (TS) is a randomized algorithm based on Bayesian ideas. Using Beta prior and considering the Bernoulli bandit problem (the rewards are either 0 or 1), TS initially assumes document $d$ to have prior $Beta(1, 1)$ on $\mu_d$ (the probability of success). At time $t$, having observed $SU_d(t)$ successes (reward = 1) and $FU_d(t)$ failures (reward = 0) in $\theta_d(t) = SU_d(t) + FU_d(t)$ selects of document $d$, the algorithm updates the distribution on $\mu_d$ as $Beta(SU_d(t)+1, FU_d(t)+1)$. The algorithm then generates independent samples from these posterior distributions of the $\mu_d$, and selects the document with the largest sample value.

\section{FA-TS}
\label{sec:crs}
To adapt the FA-TS algorithm to consider freshness document in context aware environment, we propose to compute the similarity between the present situation and each one in the situation base; if there is a situation that can be reused; the algorithm retrieves it, and then applies the TS algorithm. The proposed FA-TS algorithm is described in Algorithm \ref{alg:FA-TS} and involves for each trial $t =  1...T$ the following tasks. \textbf{Task 1:} Let $S^t$ be the current user's situation, and $PS$ the set of past situations. The system compares $S^t$ with the situations in $PS$ in order to choose the most similar $S^p$ using the $RetrieveCase()$ method. 
\textbf{Task 2:} Let $D$ be the document collection and $D^p\in D$ the set of documents recommended in situation $S^p$. After retrieving $S^p$, the system observes the user's behaviour when reading each document $d^i \in D^p$. Based on observed rewards, the algorithm chooses the document $d^p$ with the greater expected reward $r^t$ using the  $RecommendDocuments()$ method. To have the appropriate exploration at each situation, the 
$RecommendDocuments()$ method include a module $R(S^t)$ that computes the risk of the situation.
\textbf{Task 3:} The algorithm improves its document-selection strategy with the new observation $(S^t, d^t, r^t)$. The updating of the case base is done using the $Auto\_improvement()$ method.

\begin{algorithm}[H]
   \caption{The FA-TS algorithm}
\label{alg:FA-TS} 
\begin{algorithmic}[1]
 \STATE {\bfseries }\textbf{Require:}  $d \in D$ set $UP, PS, N$  
 \STATE {\bfseries }Foreach $t = 1, 2, . . . ,T$ do
 \STATE {\bfseries } $(S^p,UP^p)=RetrieveCase(S^t,PS, UP, D)$ // Retrieve the most similar case
 \STATE {\bfseries } $SelectDocuments(UP^p, S^t, S^p, D, N)$ //  Recommend N documents
 \STATE {\bfseries } \textbf{Receive a feedback $UP^t$ from the user}
 \STATE {\bfseries }$Autoimprovement(UP^p,UP^t,S^t, S^p,N)$  // Update user's profile
   \end{algorithmic}
\end{algorithm}  

\textbf{RetrieveCase():} The system compares $S^t$ with the situations in \textit{PS} in order to choose the most similar one, $S^p = argmax_{S^{i}\in PS}sim(S^t,S^i)$.
The semantic similarity metric is computed by:
\begin{equation}
 \label{eq:sim}
sim(S^t,S^i)=\sum_{\delta \in \Delta} \alpha_{\delta} sim_{\delta}( c^t_{\delta},c^i_{\delta})
\end{equation}                          
In Eq.~\ref{eq:sim}, $sim_{\delta}$ is the similarity metric related to dimension $\delta$ between two concepts $c_{\delta}^t$ and $c_{\delta}^i$, and $\Delta$ is the set of dimensions (in our case Location, Time and Social); $\alpha_{\delta}$ is the weight associated to dimension $\delta$ and it is set out by using an arithmetic mean as follows:
$\alpha_{\delta}=\frac{1}{t-1}(\sum_{k=1}^{t-1} y^{k}_{\delta})$
,where $y^{k}_{\delta}=sim_{\delta}(c^K_{\delta},c^p_{\delta})$ at trial $k \in \{ 1,...,t-1 \}$  from the $t-1$ previous recommendations, and $c^p_{\delta} \in S^p$.  
The idea here is to augment the importance of a dimension with the previously corresponding computed similarity values, reflecting the impact of the dimension when computing the most similar situation in Eq.\ref{eq:sim}.
The similarity between two concepts of a dimension $\delta$ depends on how closely $c_{\delta}^t$ and $c_{\delta}^i$ are related in the corresponding ontology. To compute $sim_{\delta}$, we use the same similarity measure as \cite{15}: 
\begin{equation}
\label{eq:simdelta}
sim_{\delta}(c_{\delta}^{t},c_{\delta}^{i})=2*\frac{depth(LCS)}{depth(c_{\delta}^t)+ depth(c_{\delta}^i)}
\end{equation}  
In Eq.~\ref{eq:simdelta}, $LCS$ is the Least Common Subsumer of $c_{\delta}^t$ and $c_{\delta}^i$, and $depth$ is the number of nodes in the path from the current node to the ontology root.
 
\textbf{SelectDocuments():} The algorithm chooses the document $d^p$ with the greatest index $P$ computed as follows:
\begin{equation}
\label{eq:fats} 
 P(d) = (1-\epsilon) \ast \theta(d,S^p)- \epsilon \ast Mr(d)
\end{equation} 
In Eq.~\ref{eq:fats}, 
$\theta(d, S^p)= SU_d(S^p,t) + FU_d(S^p,t)$. The idea here is to consider the sampling for each user's situation rather than all over the situations.

$Mr(d)$ is the strength of strangeness or the amount of experience forgotten. We apply Forgetting Curve \cite{ebbinghaus1913memory} to evaluate the freshness of a document to a user. The Forgetting Curve is shown as follows:
\begin{equation}
\label{eq:mr} 
Mr(d) = e^{-\frac{t(d)}{rsm(d)}}
\end{equation} 	                         
In Eq.~\ref{eq:mr}, $Mr$ is memory retention, $rsm$ is the relative strength of memory and $t$ is time. The least the amount of memory retention of a document is in a user's mind, the freshest is the document to the user. In our work, $rsm$ is defined as the number of times the document has been clicked and $t$ is the distance from present time to the last time the document has been clicked.

To adapt the impact of the user's memory retention to context-aware environment, we consider an $\epsilon$ that manage the weight of the $Mr$ in computing the pertinence of documents. With the assumption that more the situation is risky more the user does not forget the document related to this situation, we propose to reduce recommending fresh document according the risk of the situation. More the situation is risky less fresh document is explored. Concretely, the algorithm computes the weight of $\epsilon$, by using the situation risk level $R(S^t)$, as indicated in Eq.~\ref{eq:epsilon}.
\begin{equation}
 \label{eq:epsilon}
\epsilon = \epsilon_{max}-R(S^t)*(\epsilon_{max}-\epsilon_{min})
\end{equation}
A strict exploitation ($\epsilon$=0) leads to a non optimal documents selection strategy, this is why $R$ is multiplied by $(1-\epsilon _{min})$, where $\epsilon_{min}$ is the minimum exploration allowed in $CS$ and $\epsilon_{max}$ is the maximum exploration allowed in all situations (these metrics are fixed to $\epsilon_{max}= 0.5 \wedge \epsilon_{min}=0.05$ using an off-line simulation).

\textbf{Autoimprovement():} Depending on the similarity between the current situation $S^t$ and its most similar situation $S^p$, two scenarios are possible: (1) If sim($S^t$, $S^p$) $\neq 1$ \emph{then} $PS= PS \cup S^t \wedge UP=UP \cup UP^t$ : the current situation does not exist in the case base; the system adds this new case composed of the current situation $S^t$ and the current user preferences $UP^t$; (2) If sim($S^t$, $S^p$) $= 1 $ then $S^p= S^p \cup S^t \wedge UP^p=UP^p \cup UP^t$: the situation exists in the case base; the system updates the case having premise the situation $S^p$ with the current user preferences $UP^t$. 

\textbf{Computing the Risk Level of the Situation:}
The risk complete level \textit{$R(S^t)$} of the current situation is computed by aggregating three approaches $R^c$, $R^v$ and $R^m$ as follows:
\begin{equation}
\label{eq:rst}
R(S^t)=\sum_{j \in J}\lambda_{j}R_{j}(S^t)
\end{equation}  
In Eq.~\ref{eq:rst}, $R_{j}$ is the risk metric related to dimension $j \in J$, where $J=\{m, c, v\}$; $\lambda_{j}$ is the weight associated to dimension $j$ and it is set out using an off-line evaluation. $R_c$ compute the risk using concepts, $R_m$ compute the risk using the semantic similarity between the current situation and situations stocked in the system and $R_v$ compute the risk using the variance of the reward. The three approaches and their aggregation are described in \cite{bouneffouf2013drars}.

\section{Evaluation of FA-TS}
\label{sec:5}
In order to empirically evaluate the performance of our approach in on-line environment, we conduct our experiment with 3500 users of mobile application. We have randomly split users on five groups, and we assign to each group the mobile application with different recommendation algorithms (the algorithms are described below). Each time the user opens his software he gets 10 documents recommended by the system. 
To evaluate the impact of the risk we compare FA-TS to a variant with a fixed $\epsilon$ exploration of freshness like: \textbf{FA-TS-1: } In FA-TS, the risk is fixed to 1 ($\epsilon=0$), which means that the algorithm does not consider the freshness in its recommendation. \textbf{FA-TS-0.5:} In FA-TS, the risk is fixed to 0.5 ($\epsilon=0.5$), which means that the algorithm considers the freshness of the documents and the probability computed by the TS to recommend document. \textbf{FA-TS-0:} In FA-TS, the risk is fixed to 0 ($\epsilon=1$), which means that the algorithm considers just the freshness to recommend document (no consideration of the risk of the situation) and \textbf{TS:} The TS uses the algorithm described in \cite{TSCB} to recommend document without consideration of freshness documents.
  
\begin{figure}[!ht]
    \centering
    \centering{\includegraphics[width=0.5\columnwidth]{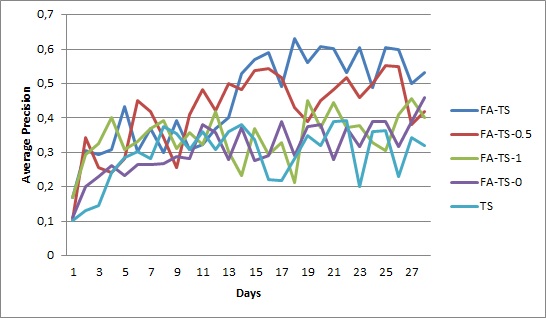}}
    \qquad
    \begin{tabular}[b]{cc}\hline
      Algorithms & AP     ~~  ATSD \\ \hline
      FA-TS      & 0,6542     1,3705 \\
      FA-TS-0.5  & 0,6187     1,3701 \\
      FA-TS-1    & 0,5450	  1,3585\\
      FA-TS-0    & 0,5109	  1,3605\\
      TS         & 0,4950	  1,3714\\ \hline
    \end{tabular}
     \label{tab:recalg}
    \caption{Average Precision on top 10 documents for each algorithm}
  \end{figure}

\textbf{Average precision on top 10 documents.} 
We compare the algorithms regarding the precision which is the number of user's clicks on the 10 recommended documents during a navigation session. The average precision (AP) is the mean of the  system's precision for all session during one day, a navigation session is the interval between the time when the user opens the mobile application and the time when he closes it. Note that we do not compute the recall because we can not know a priori all pertinent documents. In Fig.~\ref{tab:recalg}, the horizontal axis represents the day of the month and the vertical axis is the performance metric. 

We have displayed in the Table.~\ref{tab:recalg} the average number of clicks per recommendation and the average time spent on documents (ATSD) for all the 28 days. 
We have several observations regarding the different algorithms. From the Fig.~\ref{tab:recalg} we can observe that the FA-TS algorithm has effectively the best average precision during this month. We have also observed that $FA-TS-1$ gives better results than $TS$ in term of average clicks, which shows that considering the user's situation awareness in the $TS$ approach improves its result.
$FA-TS-0.5$ gives better result than $FA-TS-1$, which is explained by the consideration of the documents freshness in the TS. $FA-TS$ outperforms $FA-TS-0.5$, which shows that managing the freshness of the document according to the situation's risk gives better result than a fixed approaches. An other interesting observation is in the fact that $FA-TS-0$ outperform $TS$, which shows the impotence of considering the freshness which is not done by the $TS$.
From the Table.~\ref{tab:recalg} we can say that the ATSD does not significantly change from an algorithm to an other, which means that the exr/exp trade-off does not impact the user's time spent on documents and let us say that FA-TS gives better result on precision without reducing the quality of the recommended documents.

\section{Conclusion}
In this paper, we have studied the problem of document freshness in CARS and have proposed a new approach that considers the freshness of the document in recommendation regarding the user's situation.  
The experimental results demonstrate that considering the freshness on CARS significantly increases their performance. Moreover, this study yields to the conclusion that managing the recommendation of fresh document according to the risk of the situation gives a real add-value in recommendation performance.

\bibliographystyle{abbrv} 
\bibliography{97} 

\end{document}